%% file: ms.arxiv.tex
\newif\iftwocol

\documentclass[authoryear,preprint,12pt]{elsarticle} \twocolfalse

\usepackage{lineno}
\usepackage{amssymb}
\usepackage{textcomp}
\usepackage{wasysym} 
\usepackage[outdir=./]{epstopdf}

\newcommand{\um}{\ensuremath{\,\mu{\rm m}}}
\newcommand{\odeg}{\ensuremath{^\circ }}

\newcommand{\be}{\begin{equation}} 
\newcommand{\ee}{\end{equation}}

\input{aas_journal_abbrevs.tex}

\journal{Icarus}
\begin{document}
\begin{frontmatter}

\title{The Near-Surface Methane Humidity on Titan\\ \vspace{10pt}
	\small{\textit{Accepted for publication in Icarus}}}

\author[label1]{Juan~M.~Lora}
\ead{jlora@ucla.edu}

\author[label2]{M\'at\'e~\'Ad\'amkovics}
\ead{mate@berkeley.edu}

\address[label1]{Department of Earth, Planetary and Space Sciences, \\
University of California, Los Angeles, CA 90095, USA}
\address[label2]{Astronomy Department, University of California, Berkeley, CA 94720, USA} 

\begin{abstract}
We retrieve vertical and meridional variations of methane mole fraction in Titan's lower troposphere by
re-analyzing near-infrared ground-based observations from 17~July~2014~UT
\citep{A16}. We generate synthetic spectra using atmospheric methane profiles that do not contain
supersaturation or discontinuities to fit the observations, and thereby retrieve minimum saturation
altitudes and corresponding specific humidities in the boundary layer. We relate these in turn to
surface-level relative humidities using independent surface temperature measurements. We also
compare our results with general circulation model simulations to interpret and constrain the
relationship between humidities and surface liquids. The results show that Titan's lower troposphere
is undersaturated at latitudes south of 60$\odeg$N, consistent with a dry surface there, but
increases in humidity toward the north pole indicate appreciable surface liquid coverage. While
our observations are consistent with considerably more liquid methane existing at the north pole
than is present in observed lakes, a degeneracy between low-level methane and haze leads to
substantial uncertainty in determining the extent of the source region.
\end{abstract}

\begin{keyword}
Titan, atmosphere \sep Adaptive optics \sep Atmospheres, evolution \sep Atmospheres, structure
\end{keyword}

\end{frontmatter}

\section{Introduction}\label{s:Intro}

Like Earth, Titan experiences an active hydrologic cycle, which is responsible for the occurrence of
tropospheric methane clouds \citep{Brown2002,Griffith2005,Schaller2006a,Schaller2009}, precipitation
\citep{Turtle2011a}, and resulting erosional surface features like channels and flood plains
\citep{Elachi2005,Tomasko2005,Lopes2010}. Unlike on Earth, where the vast majority of water available to the climate system is in the ocean, Titan's observable methane is primarily in the atmosphere, which contains up to an order of magnitude more than the surface \citep{Lorenz2008}. Subsurface reservoirs are unknown, but since the atmospheric
lifetime of methane is geologically short due to photolysis \citep{Yung1984}, we only have
hypotheses \citep[e.g., subsurface clathrates;][]{Atreya2006} for the ultimate source of methane in
the climate system. As a result, determination of the atmospheric methane content, particularly in
the troposphere, is important and broadly interesting in the context of both the climate and
atmospheric evolution.

Methane thermodynamics is an important component of the energy budget and circulation of Titan's
atmosphere \citep{Mitchell2006,Mitchell2009b,Mitchell2012}, and the methane humidity is a key factor
in the development of convective cloud systems \citep{BarthRafkin2007,Griffith2008}. The
distribution of methane in the atmosphere may be diagnostic of the methane at or near the surface
\citep{Mitchell2008,Lora2015}: Because it is largely set by the availability of surface methane, the
near-surface humidity is particularly important for determining the contemporary distribution of
surface and subsurface methane, and its accessibility to the atmosphere. The {\it Huygens} Gas
Chromatograph Mass Spectrometer (GCMS) measured a near-surface relative humidity of roughly 50$\%$
near the equator \citep{Niemann2010}, but the processes that set that value are as yet uncertain
since the equatorial region is considered largely a desert \citep{Griffith2014,ML16}. Though it has
been suggested that low-latitude fluvial erosional features could be relics of a rainier past
\citep{Griffith2008}, potential paleolake basins at high latitudes \citep{Hayes2016} and
paleoclimate simulations \citep{Lora2014} suggest that surface methane cycles between the poles on
geologic timescales, rarely remaining at low latitudes. Yet there is evidence for present surface
liquids at low latitudes \citep{Griffith2012}, so a fully consistent picture of Titan's hydrologic
cycle remains elusive.

While the low latitudes are dry, Titan's poles support lakes of liquid hydrocarbons
\citep{Stofan2007,Hayes2008,Turtle2009}, which may be connected hydraulically over regional scales
by subsurface ``alkanofers'' \citep{Hayes2016}. These polar regions are also topographically
depressed with respect to other latitudes, and relatively devoid of craters, which suggests the
possibility of extensive high-latitude wetlands \citep{Neish2014}. Furthermore, surface temperature
measurements from {\it Cassini} Composite Infrared Spectrometer (CIRS) suggest that, as springtime
advances, the north pole has warmed more slowly than expected from a dry porous regolith
\citep{Jennings2016}. And despite the northern lakes occupying only about 10$\%$ of the polar
surface \citep{Hayes2011}, no zonal temperature contrasts have been detected, which is interpreted
as further indication of moist ground \citep{Jennings2016}. Such circumstantial evidence raises the
question: Do the lakes represent the entire source region of methane that enters the atmosphere, or
is there a larger geographical region with moisture seeping out of the regolith?

Initial studies of the methane cycle in general circulation models of Titan's atmosphere assumed an
infinite supply of surface methane globally available to the atmosphere
\citep{Rannou2006,Mitchell2006}, but more recent work has shown that the atmospheric circulation
transports methane to the poles, where it is cold-trapped
\citep{Mitchell2008,Schneider2012,Lora2015,LoraMitchell2015}, and the simulations that are most
consistent with various observations are those where the rest of the surface is largely dry
\citep{Lora2015}. In short, the interactions of surface liquids with the atmosphere, and subsurface
processes affecting these liquids, are only roughly parameterized in models, and very few
observations are available as benchmarks. An understanding of the distribution of low-level humidity
on Titan would shed some light on these issues, but measurements of the variation of humidity with
latitude have been limited and considerably hindered by Titan's opaque and nearly-saturated
atmosphere \citep{Anderson2008,Penteado2010b}.

Recently, \citet{A16} used high spectral resolution observations in the near-IR---with complementary
lower-resolution and wider bandpass measurements to constrain spatial variations in haze
opacity---to measure the meridional variation of tropospheric methane. Their simultaneous
ground-based observations from two facilities were unable to break a degeneracy between the roughly
anti-correlated methane (absorption) and haze (scattering) opacities near the surface, leading to
ambiguity in determining whether the methane or haze near the surface was variable. Simulated images
of {\it Cassini} VIMS observations of both northern and southern hemispheres suggest that a
spatially uniform near-surface haze is more likely, and circulation models with haze microphysics
support this interpretation, since they predict that haze near the surface should not have strong
spatial variations \citep{Larson2014}. A uniform surface haze means that the observed opacity
variations near the surface are indeed due to changes in the methane content.

Our motivation here is to measure and interpret the distribution of near-surface methane in Titan's
atmosphere, specifically by considering both the vertical and meridional variation in atmospheric
methane.
\citet{A16} implemented an ad hoc scale factor for the methane profile in the troposphere,
which lead to supersaturated conditions in retrieved profiles with relatively high column methane
content. Here we invoke a realistic variation in the vertical mole fraction of methane, which does not allow
supersaturation, in order to interpret high-resolution near-IR spectra. We describe our
observational and radiative methods in Section~\ref{s:obs}, and the models used for interpreting
these data in Section~\ref{s:TAM}. Results are presented in Section~\ref{s:results}, and we discuss
how our methods leverage the increased sensitivity to opacity variations at higher altitudes and
connect them in a physical way to near-surface humidities, as well as how assumptions about the
surface temperature are important for retrieving relative humidities. We describe how well our
observations constrain the distribution of methane on the surface in Section~\ref{s:discussion}.

\section{Methods}\label{s:methods}
\subsection{Observations and Simulated Spectra}\label{s:obs}

The Near-InfraRed SPectrometer \citep{McLean1998} with Adaptive Optics (NIRSPAO) was used at W.~M.
Keck Observatory on 17~July~2014~UT to observe Titan with a spectral resolving power of
$R\approx25,000$ and a spatial sampling  of 0.018"/pixel along the slit. A single north-to-south
position along the central meridian was integrated for 45 min. We analyze spectra from one echelle
order centered near 1.55 \um. Additional details of these observations, including the data reduction
and calibration with supporting datasets, are described in \citet{A16}.

Synthetic spectra are generated by defining 20 atmospheric layers, with properties that are
determined primarily by measurements made with instruments on the {\it Huygens} probe (HASI and GCMS instruments; see below), including the
temperature, pressure, methane mole fraction, and haze structure. These {\it in situ} measurements are used to
determine the gas and scattering opacity in each layer of the model. The discrete ordinate method
radiative transfer \citep[DISORT;][]{Stamnes1988} is implemented in Python (PyDISORT) and used to
solve the radiative transfer through the model atmosphere and simulate the observed flux. The
methane in the lower atmosphere and haze in the upper atmosphere are assumed to vary with
latitude \citep{A16}; however, here we consider a more physically realistic vertical (altitude)
structure for the methane gas, which is used to determine the methane relative humidity, $RH$, at
the surface.

The vertical profile of methane measured by the GCMS experiment on
the {\it Huygens} probe follows the saturation curve from roughly 40 km down to  6 km altitude,
below which the methane mole fraction is constant at 0.057 \citep{Niemann2010}. In order to explain
its meridional variation, \citet{A16} consider a methane profile that scales the
tropospheric methane mole fraction below 35\,km altitude. While straightforward, making such an
assumption has limitations. For example, in order to accommodate an increase in methane opacity,
this assumption leads to supersaturation at altitudes above 6 km. It is also unclear if the changes
in the synthetic spectra from such a variation are most sensitive to higher or lower altitude
methane mole fraction changes. While the number density of methane is largest near the surface, and therefore
should contribute the greatest amount to the total column, there are wavelength windows that are
insensitive to the surface. These windows become wider in wavelength near the limb, where the slant
paths through the atmosphere and therefore the total opacity are larger. Lastly, scaling the methane
mole fraction leads to a discontinuity at 35 km altitude, the level used to identify the top of the
troposphere.

Here we avoid the supersaturation of methane and discontinuities in the retrieved methane mole
fraction by considering profiles that have a characteristic minimum altitude of saturation, $z_s$.
We consider $z_s$ a free parameter for fitting synthetic spectra to the observations. If a greater
methane opacity is required to fit the observations, we increase the column of methane by decreasing
the saturation altitude, rather than scaling the methane mole fraction at all altitudes. Conversely,
in order to decrease the column of methane, we increase $z_s$. The methane saturation vapor pressure
at temperature $T$ is given by the Clausius-Clapeyron relation over a binary mixture with saturation vapor pressure 106 mbar at 90.7 K \citep{Thompson92,Lora2015}:
\be
\label{e:p_s}
p_{s}(T) = X_{\rm CH4}*106.*\exp(L_v/R_d*((1.0/90.7)-(1.0/T))),
\ee
where $L_v$ and $R_d$ are the latent heat of evaporation (assumed constant) and gas constant for methane, respectively. $X_{\rm CH4}=0.9$ is the fraction of methane in a methane-nitrogen mixture that best fits the GCMS data. We use the
temperature profile measured by HASI on the {\it Huygens} probe to identify the temperature at a
given altitude, $T(z)$ \citep{Fulchignoni2005}. Variations in the vertical profiles of methane are
illustrated in Fig.~\ref{f:methane_profiles}. These profiles provide a more physically plausible
variation in the total methane column than those used by \citet{A16}.

An example of a spectrum from a single pixel along the spectrometer slit is shown together with a
best fit synthetic spectrum in Fig.~\ref{f:spectra}. In this case, increasing or decreasing $z_s$
leads to residuals that are in excess of estimates of the observational uncertainty. We determine
the best value of $z_s$ for each latitude (i.e., spatial pixel along the slit) by using a
brute-force approach where we set $z_s$ to the center of each layer of the model between 0 and 20 km
altitude, generate synthetic spectra for each case, and calculate the mean squared residual,
$\widehat{\chi^2}(z_s)$, between each synthetic spectrum and the observation. This process is repeated for
spectra from each pixel along the slit.

%-----------------------------------------------------------------------------------------
%SINGLE COLUMN FIGURE
\iftwocol
	\begin{figure*} \includegraphics[]{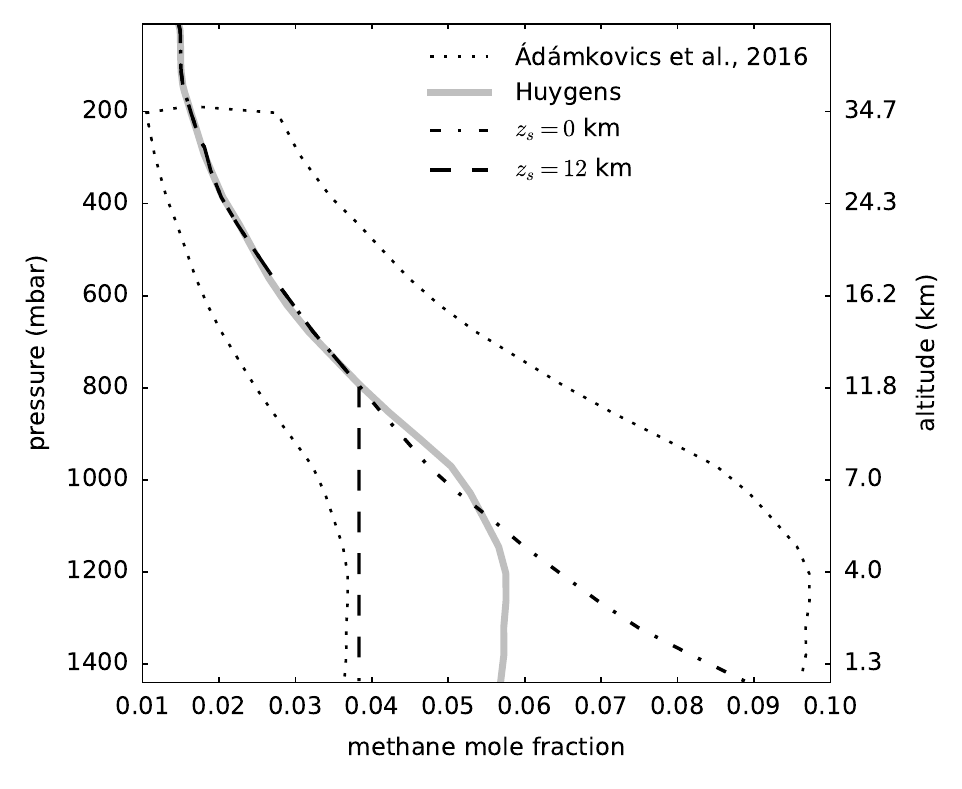} 
\else
	\begin{figure} \includegraphics[]{methane_profiles.pdf}
\fi
	\begin{center}
	\caption{\label{f:methane_profiles}
    The methane mole fraction vertical profiles used in \citet{A16} for changes to the total methane
    humidity (dotted) scale the {\it Huygens} profile (gray) below $\sim$200 mbar. This scaling
    causes supersaturation and an unphysical discontinuity. These profiles are compared with
    profiles that are defined by a characteristic minimum saturation altitude, $z_s$. Two examples
    are illustrated: saturation to the surface, $z_s$=0 (dash-dot), and a minimum saturation
    altitude of $z_s$=11 km altitude (dashed). The methane mole fraction is assumed to be constant
    below $z_s$, similar to the {\it Huygens} profile near the surface.}
	\end{center} 
\iftwocol
	\end{figure*}
\else	
	\end{figure}
\fi
%-----------------------------------------------------------------------------------------

%-----------------------------------------------------------------------------------------
%DOUBLE COLUMN FIGURE
\iftwocol
	\begin{figure*} \includegraphics[]{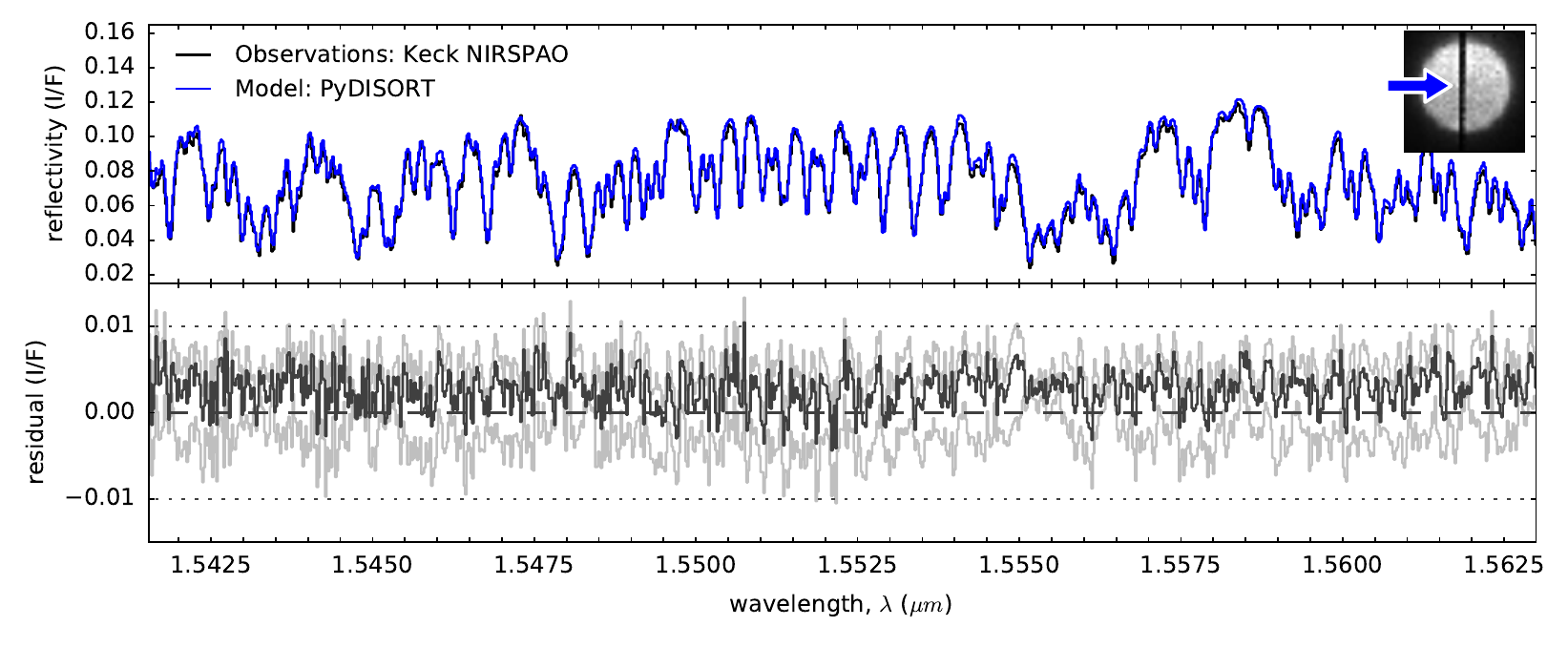} 
\else
	\begin{figure} \includegraphics[]{example_spectra.pdf}
\fi
	\begin{center}
	\caption{\label{f:spectra}
	The spectrum from one NIRSPAO pixel (top panel; black), at the location specified by the blue
	arrow in the slit-viewing camera image (inset), is compared with the radiative transfer model
	spectrum (blue). The residuals are plotted in the bottom panel, with the ordinate scale
	magnified relative to the top panel. Residuals from radiative transfer models with minimum
	saturation altitudes that are either higher or lower in the atmosphere (gray) illustrate the
	spectral sensitivity to this parameter. }
	\end{center} 
\iftwocol
	\end{figure*}
\else	
	\end{figure}
\fi
%-----------------------------------------------------------------------------------------

\subsection{Titan Atmospheric Model}\label{s:TAM}

We interpret the retrieved profiles in the context of Titan's climate and general circulation by
comparison to simulations with the Titan Atmospheric Model (TAM), a three-dimensional general
circulation model (GCM) that includes parameterizations for nongray radiative transfer, Titan's
hydrologic cycle, surface, and boundary layer \citep{Lora2015}. A simple ``bucket" model, wherein
surface methane at each gridbox is replenished by precipitation and is available to evaporate into
the atmosphere as long as it is above a minimum value, is used to represent surface/subsurface
hydrology; no horizontal surface/subsurface transport occurs. The simulations here use a 32-layer
(L32) atmosphere \citep{Lora2015} at T42 resolution (approximately 2.8$\odeg$ resolution).

As a baseline and for comparison of our results to previous GCM simulations that either prescribed
the surface humidity or assumed an infinite surface methane reservoir
\citep{Mitchell2006,Rannou2006,Tokano2009}, we consider an ``aquaplanet" scenario in which
practically inexhaustible surface methane liquid is available globally.  We also use results from
previous (T21) simulations \citep{Lora2015} in which only the locations of the largest observed
lakes contain large amounts of surface methane, but other regions are allowed to moisten and dry
self-consistently. Finally, we employ simulations following the ``wetlands" scenario used in
\citet{LoraMitchell2015} and motivated by \citet{Neish2014}, where ample surface liquid is
initialized poleward of 60$\odeg$ in both hemispheres. Under this configuration, we use a range of
simulations varying the calculation of methane saturation vapor pressure to facilitate a more
exhaustive comparison between simulation results and the observation analysis outlined above. These
include parameterizations that include both liquid-vapor and ice-vapor phase transitions
\citep[i.e., simulations in][]{LoraMitchell2015}, or that assume only the liquid-vapor phase
transition over liquid (Eq.~\ref{e:p_s}) with a range of 80$\%$--100$\%$ methane content ($X_{\rm
CH4}=0.8$--1.0).

Fig.~\ref{f:hum_model} shows the zonal-mean surface-level specific humidities from the various TAM
simulations during late northern spring, corresponding to the time of the observations. At low
latitudes, the ``wetlands" and ``observed lakes'' cases coincide in simulating lower specific
humidities than at high latitudes, though in the latter case the contrast is considerably smaller as
a result of the sparse coverage of polar liquids. In contrast, the ``aquaplanet'' case predicts the
highest specific humidities near the equator, reflecting the global liquid coverage and peak
annual-mean evaporation (responding to insolation) occurring at low latitudes. For comparison, the
GCMS measurement (Fig.~\ref{f:methane_profiles}) corresponds to a surface value of specific humidity
of roughly 0.032 kg kg$^{-1}$ at 10$\odeg$S, though at a different season from our observations.

The corresponding surface-level relative humidities from the simulations (right panel of
Fig.~\ref{f:hum_model}) further accentuate the differences between the various simulated cases. The
range of ``wetland'' cases bridges the high- and low-latitude predictions from the ``aquaplanet''
and ``observed lakes'' configurations with saturated conditions near the poles and much lower
humidities at low latitudes, while the latter simulations predict consistently saturated
($RH\approx100\%$) or undersaturated ($RH<75\%$) conditions, respectively. The qualitatively
different curves in Fig.~\ref{f:hum_model} illustrate the ability of near-surface humidity
measurements to distinguish between possible scenarios of the surface methane distribution on Titan.

%-----------------------------------------------------------------------------------------
%1.5-COLUMN FIGURE
\iftwocol
	\begin{figure*} \includegraphics[]{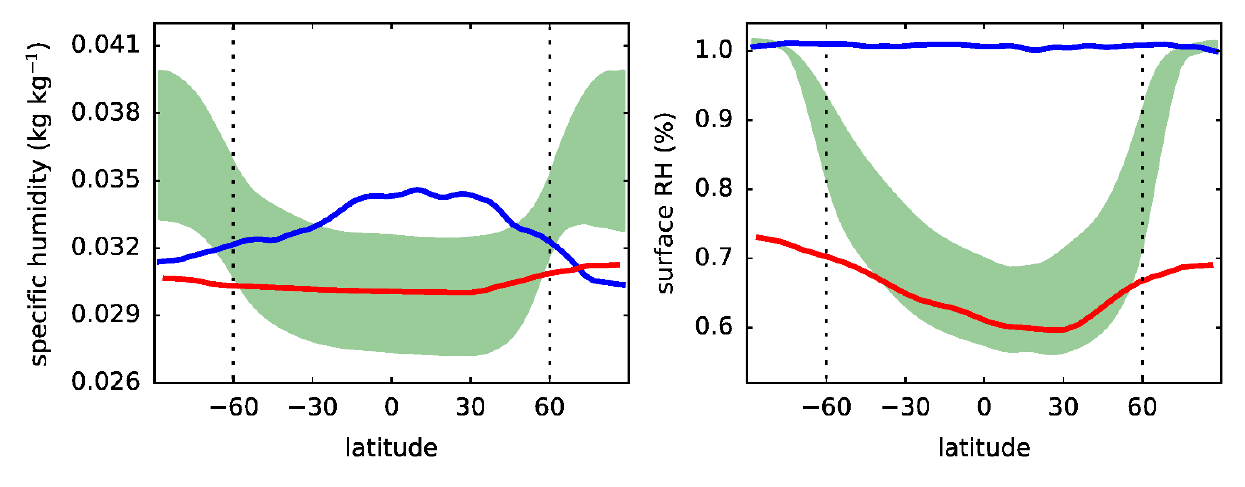} 
\else
	\begin{figure} \includegraphics[]{humidities_models.pdf}
\fi
	\begin{center}
	\caption{\label{f:hum_model}
	The simulated surface specific humidities at the season of the observations (left). The green
	area shows the range of results for ``wetlands'' simulations with various saturation
	parameterizations, including those from \citet{LoraMitchell2015}. Blue and red curves show
	simulations under ``aquaplanet'' and ``observed'' lake configurations \citep{Lora2015},
	respectively. The corresponding surface-level relative humidities from the various TAM
	simulations (right) for the season of the observations. Colors are the same as in the left
	panel. Dotted vertical lines mark the edges of ``wetlands.'' }
	\end{center} 
\iftwocol
	\end{figure*}
\else	
	\end{figure}
\fi
%-----------------------------------------------------------------------------------------

Relative humidity depends strongly on temperature as well as specific humidity. The surface-level
temperatures from the simulations at the season of observations are shown in Fig.~\ref{f:temps},
along with minimum surface temperatures derived from {\it Cassini} CIRS observations between 2013
and 2014 \citep{Jennings2016}. As discussed in \citet{Lora2015}, the considerably smaller
equator-to-pole contrast in the ``aquaplanet'' case compared to the observed temperatures refutes
the possibility of global surface liquid existing on Titan. On the other hand, the rest of the
simulations largely succeed in reproducing the temperature observations barring relatively minor
discrepancies, lending confidence to the usefulness of comparison between retrieved and simulated
relative humidities in these cases.

%-----------------------------------------------------------------------------------------
%SINGLE COLUMN FIGURE
\iftwocol
    \begin{figure*} \includegraphics[]{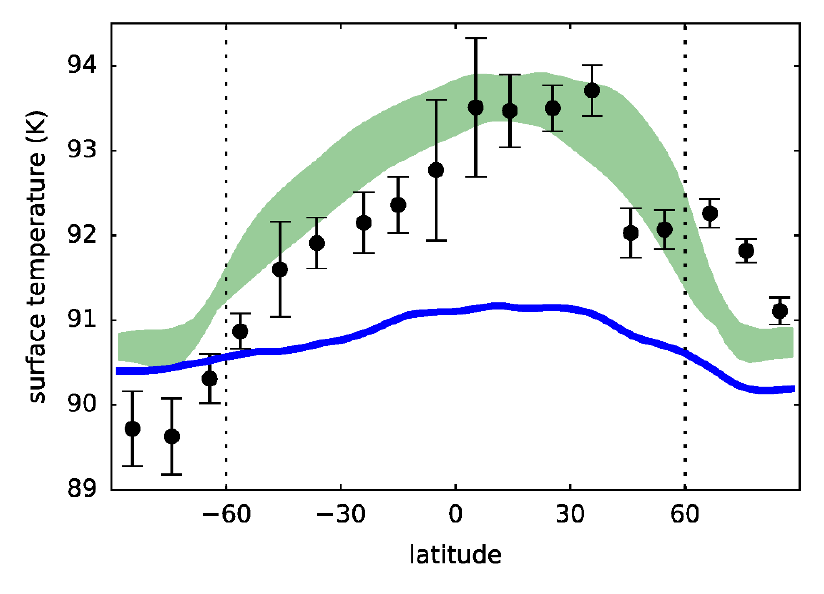} 
\else
    \begin{figure} \includegraphics[]{temperatures.pdf}
\fi
    \begin{center}
    \caption{\label{f:temps}
    The zonal-mean surface temperatures corresponding to the time of the observations. Green area
    shows the range of results for both ``wetlands'' and ``observed lakes'' simulations; blue curve
    shows the ``aquaplanet'' simulation. Dotted vertical lines mark the edges of ``wetlands.'' Black
    points with error bars show surface brightness temperatures measured between April 2013 and
    September 2014 by {\it Cassini} CIRS \citep{Jennings2016}. }
    \end{center} 
\iftwocol
    \end{figure*}
\else   
    \end{figure}
\fi
%-----------------------------------------------------------------------------------------

\section{Results}\label{s:results}

Retrievals of the minimum saturation altitude $z_s$ for all latitudes observed along the slit
recover the meridional variation in methane mole fraction reported by \citet{A16}. Fig.~\ref{f:z_s} illustrates that the trend of increasing methane toward the southern latitudes and north pole corresponds to a decreasing minimum saturation altitude in these regions. Retrievals from the few pixels closest to
the poles are consistent with saturated profiles down to the surface; however, the large slant path
through the atmosphere near the observed limb of Titan means that the sensitivity to the
near-surface is reduced as the total opacity from both scattering and gas absorption increase.

%-----------------------------------------------------------------------------------------
%SINGLE COLUMN FIGURE
\iftwocol
    \begin{figure*} \includegraphics[]{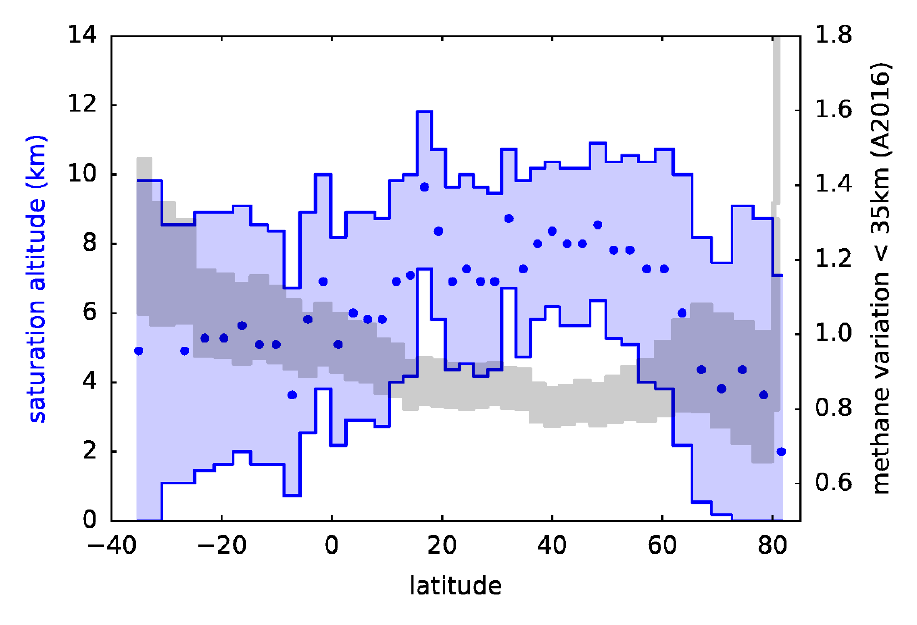} 
\else
    \begin{figure} \includegraphics[]{saturation_altitude.pdf}
\fi
    \begin{center}
    \caption{\label{f:z_s}
    The minimum saturation altitude $z_s$ determined by fitting spectra observed on
    17~July~2014~UT at each location along the slit (blue) compared to the methane variation
    presented in \citet{A16} (gray). Regions that were reported to have increased (decreased)
    tropospheric methane correspond to lower (higher) saturation altitudes. The sub-observer
    longitude is 291\odeg W.}
    \end{center} 
\iftwocol
    \end{figure*}
\else   
    \end{figure}
\fi
%-----------------------------------------------------------------------------------------

As described in \citet{A16} and discussed in the following section, there is a degeneracy between
the opacity due to methane and haze near the surface. We consider several scenarios for the
retrievals, which are outlined in the Appendix. We find that in the cases where the low-altitude
haze opacity is allowed to vary, the degeneracy in the opacities removes meridional variations in
the retrieved saturation altitudes. However, simulated images of {\it Cassini} VIMS observations in
both northern and southern hemispheres suggest that a uniform near-surface haze is more likely
\citep{A16}. Moreover, predictions from circulation models that include haze microphysics
\citep{Rannou2004,Larson2014} indicate little meridional variability in the haze in the lower
troposphere. Nevertheless, in the following we also consider the retrievals with variable low-level
haze.

Assuming a constant tropospheric haze opacity, the retrieved saturation altitudes in the southern
hemisphere are roughly constant, with the minimum $\widehat\chi^2(z_s)$ suggesting values of between 5 and 6
km, in agreement with the value inferred from {\it in situ} observations \citep{Niemann2010}. These
increase to about 8 km in the low latitude northern hemisphere, where they are also roughly constant
between 15$\odeg$ and 50$\odeg$N. Poleward of this, there is a pronounced turnover in the retrievals
as the saturation altitudes drop from $\sim$7 to $\sim$4 km around 60$\odeg$N, and the lowest values
occur nearest the north pole. In the cases where the tropospheric haze opacity is assumed to be
variable (Fig.~\ref{f:hum_obs}), the preferred saturation altitudes are essentially constant at 6 km
at all latitudes, but with increasing uncertainties toward the north that are consistent with both
increasing and decreasing values, including saturation at the surface near the north pole. Similar
uncertainties are not present toward the southern hemisphere, and in this case the values are
constrained to within the same range as in the low latitude northern hemisphere.

%-----------------------------------------------------------------------------------------
%DOUBLE COLUMN FIGURE
\iftwocol
    \begin{figure*} \includegraphics[]{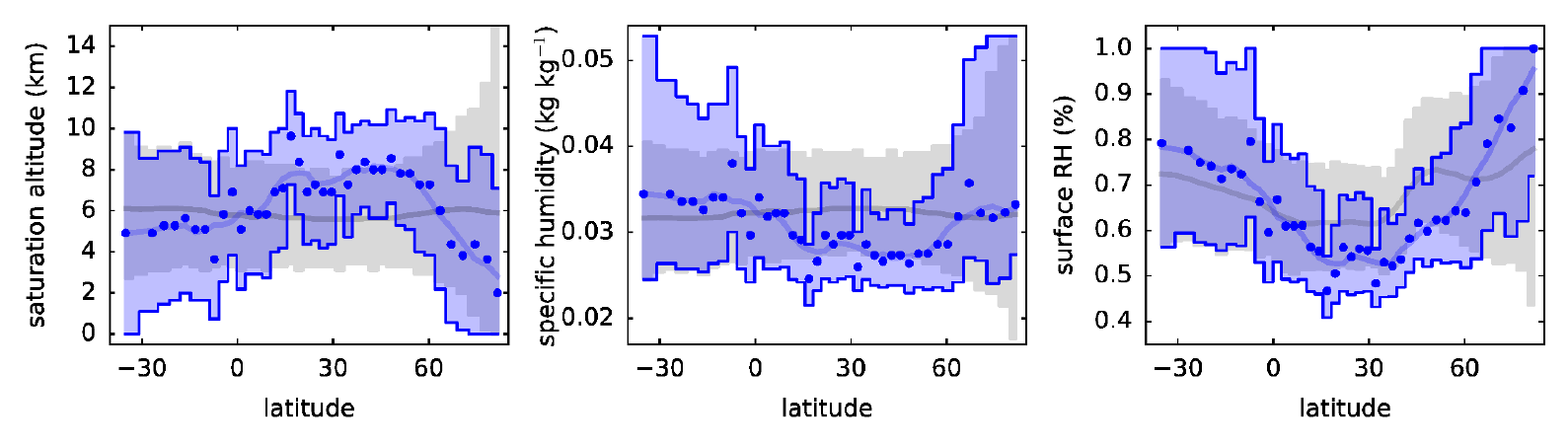} 
\else
    \begin{figure} \includegraphics[]{humidities_obs.pdf}
\fi
    \begin{center}
    \caption{\label{f:hum_obs}
   The meridional variation of minimum saturation altitudes (left), corresponding specific
    humidities (middle), and associated surface-level relative humidities (right) retrieved from the
    observations. Retrievals assuming a constant haze opacity (see text) are shown in blue, while
    retrievals assuming a variable haze opacity are shown in gray. Curves of the running mean of the
    retrieved values are shown for ease of comparison. }
    \end{center} 
\iftwocol
    \end{figure*}
\else   
    \end{figure}
\fi
%-----------------------------------------------------------------------------------------

The variation with latitude of the specific humidity associated with the retrieved minimum
saturation altitudes is consistent with higher low-level moisture content in the southern hemisphere
as well as the north polar region, with a minimum at the low- to mid-latitudes of the northern
hemisphere in the case of constant haze opacity (Fig.~\ref{f:hum_obs}, center panel). With the
possible exception of a few locations, we retrieve values for the surface specific humidity in
these regions that are the same (within error) to that measured by the GCMS on {\em Huygens}
at 10\odeg S. The same is true of specific humidities retrieved assuming a variable haze, in which
case the values are more constant with latitude but again are consistent with considerable
variations at high northern latitudes, up to a $\sim$60$\%$ enhancement near the pole.

Taking the specific humidity to be nearly constant in the sub-saturated layer below the altitude of
saturation (see Fig.~\ref{f:methane_profiles}), the above retrievals can be combined with
independent observational estimates of the surface temperatures (Fig.~\ref{f:temps}) to yield the
surface-level relative humidities. In both haze opacity cases, the resulting relative humidity
variations with latitude occur in a wide ``U'' shape that mostly results from the variation of
surface temperatures. Nevertheless, the values from northern mid-latitudes to the south are mostly
consistent with a sub-saturated lower troposphere, with the lowest relative humidities at the
latitudes of peak temperatures (in agreement with the simulations, Fig.~\ref{f:hum_model}). In the
southern hemisphere, the two retrievals are consistent with each other, but the lowest relative
humidities are roughly 10$\%$ lower in the case of constant haze opacity in the region of highest
temperatures, roughly 15$\odeg$--45$\odeg$N. This case also exhibits a considerable variation toward
the north pole, with preferred values surpassing 80$\%$ relative humidity at latitudes north of
60$\odeg$N. In both cases, the profiles are consistent with low-level saturation in the northern
polar regions.

Fig.~\ref{f:hum_compare} compares the retrieved surface-level specific and relative humidities to
those simulated by TAM. Despite the considerable uncertainties, the meridional variations of the
specific humidity from the observations are compatible with the ``wetlands'' simulations, with the
lowest values occurring at northern mid-latitudes (during the northern spring season), and
increasing humidities toward the north pole. The associated relative humidities display
similar---though more pronounced---meridional trends, with considerable low-level sub-saturation at
low- and mid-latitudes, and increasing saturation toward the north pole.

%-----------------------------------------------------------------------------------------
%1.5-COLUMN FIGURE
\iftwocol
    \begin{figure*} \includegraphics[]{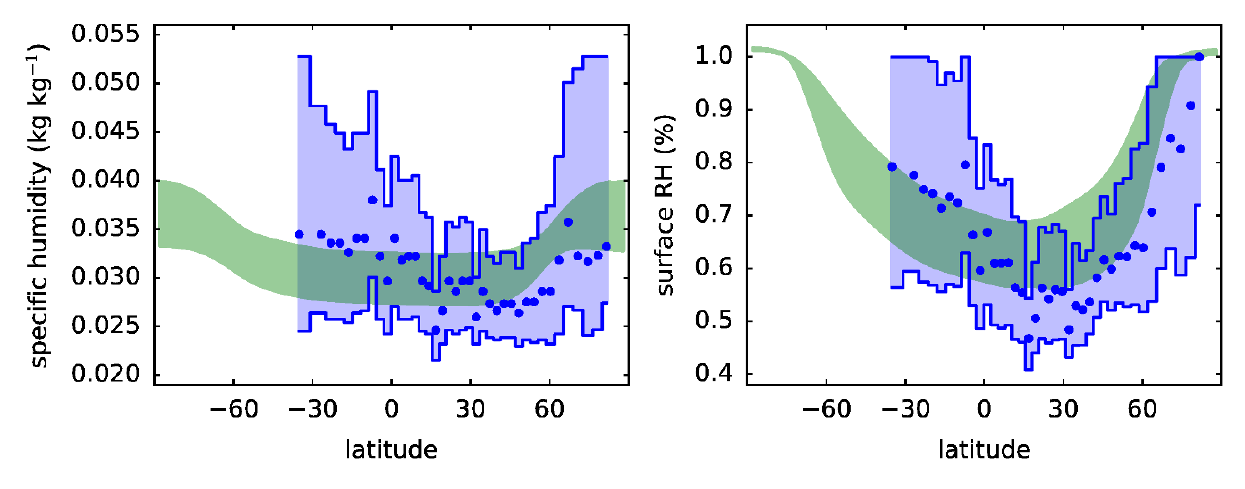} 
\else
    \begin{figure} \includegraphics[]{humidities_comparison.pdf}
\fi
    \begin{center}
    \caption{\label{f:hum_compare}
   Comparison of the retrieved and simulated specific (left) and relative (right) humidities near
	 the surface. Blue points with uncertainty envelopes are the same as in Fig.~\ref{f:hum_obs},
	 while the green areas are the simulation results as in Fig.~\ref{f:hum_model}. }
    \end{center} 
\iftwocol
    \end{figure*}
\else   
    \end{figure}
\fi
%-----------------------------------------------------------------------------------------

\section{Discussion}\label{s:discussion}
Comparison of the retrieved to simulated near-surface methane humidities (cf.
Figs.~\ref{f:hum_model} and \ref{f:hum_obs}) convincingly eliminates the ``aquaplanet'' scenario as
a plausible configuration for Titan's surface liquids, which calls into question the details of
hydrologic cycles simulated by models that prescribe excessive surface methane at low latitudes.
However, while the high-latitude ``wetlands'' distribution is compatible with the retrievals
(Fig.~\ref{f:hum_compare}), we cannot conclusively rule out the possibility that Titan's surface
methane available to the atmosphere exists exclusively in the observed northern lakes; the
``observed lakes" simulation falls in the lower values of humidity within the retrievals'
uncertainties.

With the constraints provided by our observations and simulations, it is straightforward to
interpret the increase in the near-surface humidity at the north pole as a consequence of the
existence of surface liquids there, whether or not they are more extensive than the observed lakes.
This implies that such surface methane humidifies the regional boundary layer, which in turn
transports its moisture to lower latitudes as a result of atmospheric dynamics
\citep{LoraMitchell2015,ML16}. As a corollary, the northern polar near surface should increase its
methane humidity as the seasons progress toward summer solstice and evaporation from the lake
surfaces intensifies, likely triggering convective cloud activity if the overlying troposphere is
sufficiently unstable.

Conversely, it is difficult to reconcile the apparent retrieved increase in near-surface humidity
toward the southern hemisphere (in the constant haze opacity case) with any of the simulations.
\citet{A16} interpret the increase in column methane toward the south pole, approaching winter at
the time of the observations, as possibly arising from evaporation from moist ground, as Ontario
Lacus is a vastly insufficient source of methane and such an increase implies distinct source
regions of methane vapor in the two hemispheres. But even the TAM ``wetlands'' simulation, with
surface methane covering the entire south pole, is only marginally consistent with the retrieved
specific humidities toward the south (Fig.~\ref{f:hum_compare}). Two possibilities emerge as
solutions to this conundrum: Either much more of the southern hemisphere is covered in liquid
(possibly down to the low latitudes), or the assumption of constant low-level haze opacity begins to
fail in the southern hemisphere. An inherent problem with the former option is that such a liquid
distribution would push the system closer to the ``aquaplanet'' configuration, which is incompatible
with observations of Titan's surface \citep{Stofan2007,Lorenz2006a,ML16}, so we are left to conclude that the observational uncertainties toward
the south preclude a determination of the meridional variation of low-level atmospheric methane
there.

As a result, a conservative interpretation is that Titan's near-surface meridional humidity
variations roughly fall in the areas of overlap of the uncertainty envelopes of our two end-member
retrievals. As such, low-latitude variations are difficult to make out, but the increase in humidity
toward the north pole is a more robust feature (though with still substantial uncertainties). The
same general trend is true of the ``wetlands'' simulations, where subtle low-latitude variations of
humidity give way to large increases toward the polar regions.

From late 2006 through early 2007, \citet{Brown2009b} measured low-altitude fog near the south pole,
indicating a saturated atmosphere above the surface. There did not seem to be a correlation with the location of Ontario Lacus, and the meteorological conditions giving rise to the observed
phenomenon remain unknown. If we speculate that the existence and location of fog, like
higher-altitude clouds, are seasonally variable, and further assume that the meteorology of the
north and south poles are similar, then saturated conditions near the surface at the north pole may
be consistent with the formation of fog there. Predictions from circulation models suggest that
analogs of the large clouds seen during southern summer should already be occuring near the north
pole; it is unclear if the current absence of prominent convective clouds near the north pole is
related to the possibility of saturated conditions and the stability of the troposphere.

Future near-IR observations used to measure the methane content near the surface will continue to
face the challenge of differentiating between the opacity due to methane absorption and that due to
haze scattering. This degeneracy exists even though the methane spectrum in the near-IR is highly
variable with wavelength, while the haze opacity is not. The difficulty arises from the fact that in
the spectral regions where methane opacity is low (and the reflectivity is high)---wavelengths that
are sensitive to atmospheric opacity down to the surface---an increase (decrease) in the methane
mole fraction leads to changes in the reflectivity spectrum that are similar to a decrease
(increase) in the haze opacity. The many weak, densely-spaced, and pressure-broadened methane line
wings contribute to a pseudo-continuum in the methane line opacity.

While very high S/N observations may be able to resolve the differences in line wing shapes due to
methane variation from those due to haze variation, it is unclear whether systematic instrumental
uncertainties will limit the S/N that can be achieved. Spectra at longer wavelengths, approaching
the mid-IR, have less densely spaced methane lines, which suggests that there may be a better chance
of resolving line profiles that are sensitive to the methane content near the surface. However,
telluric thermal emission makes ground-based observation of Titan at these wavelengths prohibitively
difficult. Combining simultaneous observations in the visible, near- and mid-IR, at both low and
high scattering phase angles could probe both the methane and haze opacity in a way that might break
the degeneracy for methane retrievals.

\break

\section*{Acknowledgements} 
The authors thank D. Jennings for providing surface temperature measurements, and J. Mitchell for
important discussions. M.{\'A}. was supported by NASA grant NNX14AG82G.

\break

\appendix
\setcounter{figure}{0}

\section{Spectral fitting methodology}

Since $z_s$ has discrete values for each layer in our radiative transfer model, we follow a
brute force approach for optimization of this model parameter. For each particular observing
geometry that corresponds to a pixel in the slit, we set $z_s$ to each layer of the model and
determine a goodness-of-fit parameter \citep[e.g., as in][]{A10} that is the ratio of the mean
absolute deviation of the residual in the spectrum for each channel, $\lambda$, relative to the
observational uncertainty,
\begin{equation}
\widehat{\chi^2}(z_s) = \frac{1}{n} \sum_{i=1}^n 
\frac{|I_{\rm obs}(\lambda) - I_{\rm calc}(\lambda; z_s)|}{\sigma_{\rm obs}}.
\end{equation}
A detailed analysis and determination of the observational uncertainty, $\sigma_{obs}=0.002\,I/F$,
is presented in \citet{AM16}. Values of $\widehat{\chi^2}(z_s)$ are calculated for $z_s$ set to the
altitude at the middle of each model layer from the surface up to 18\,km. The discrete values of
$\widehat{\chi^2}(z_s)$ are then fit with a second-order polynomial. The minimum in the polynomial
is the optimized value for $z_s$ for that pixel, and the 1$\sigma$ uncertainty in the optimization
is evaluated where the fit to $\widehat{\chi^2}(z_s) = 1$. The discrete values and fitting are
illustrated for each spectrum in Figure~\ref{f:chi2}.

The optimization near the limb has a greater uncertainty than near the center of the disk. Near the
limb, where the slant path through the atmosphere is largest and where spectral sensitivity to the
surface is smallest, the optimization of $z_s$ is asymmetric. In these regions the lower limit to the
uncertainty in $z_s$ is overestimated by the polynomial fit to the discrete values of
$\widehat{\chi^2}(z_s)$. In these regions a total methane column that is in excess of saturation
down to the surface could still be consistent with observations. This means that the spectra from
near the limb could also be consistent with a warmer temperature profile than used in the model, or supersaturated conditions.

%-----------------------------------------------------------------------------------------
%DOUBLE COLUMN FIGURE
\iftwocol
    \begin{figure*} \includegraphics[width=1\textwidth]{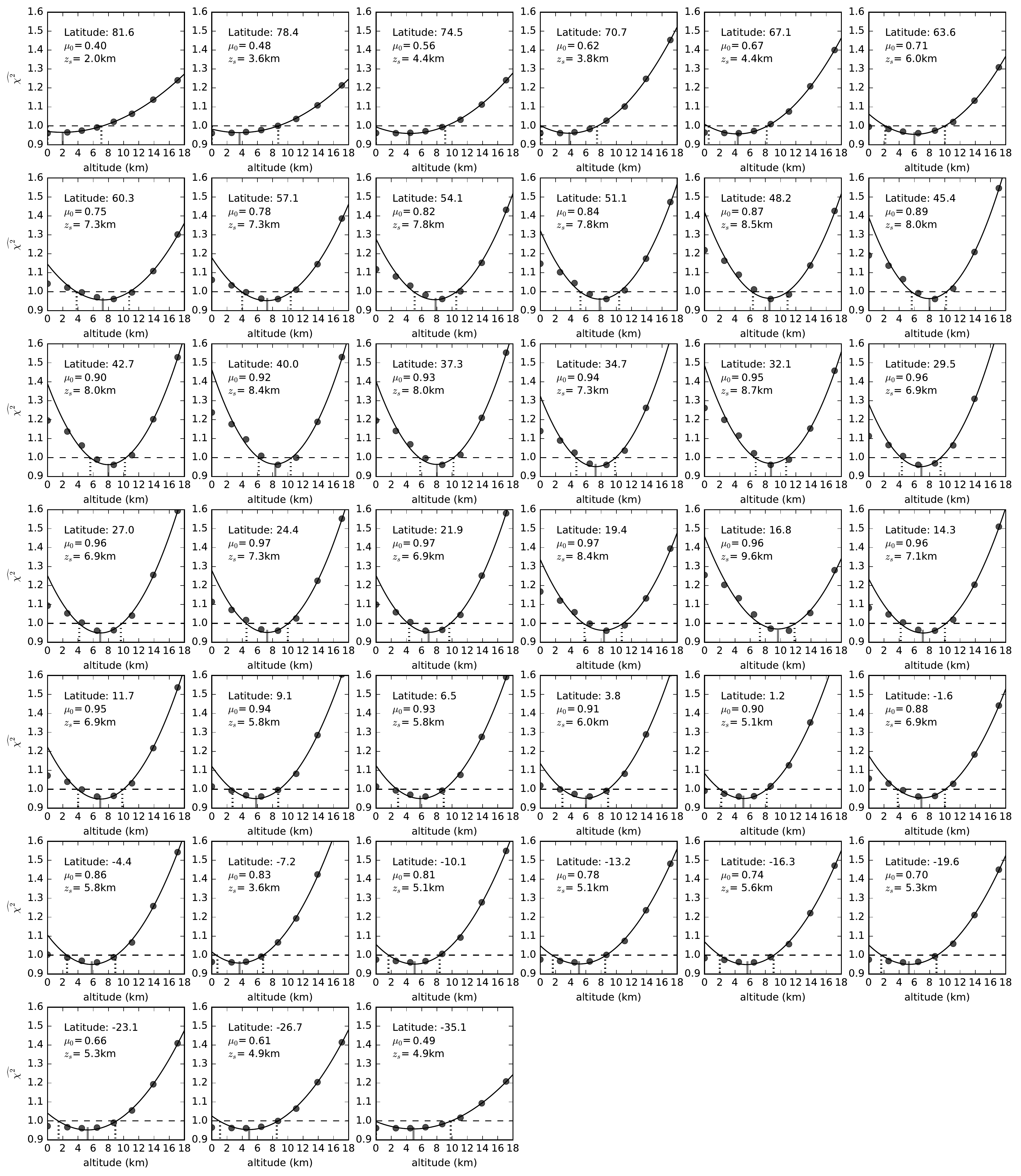} 
\else
    \begin{figure*} \includegraphics[width=1\textwidth]{chi2.pdf}
\fi
    \begin{center}
    \caption{\label{f:chi2} The optimization for $z_s$ presented in 
    Figure~\ref{f:z_s}, via the quality-of-fit parameter $\widehat{\chi^2}$, shown for each
    spectrum. The corresponding latitude, incidence angle $\mu_0=\cos\theta_0$, and best fit $z_s$
    are given in each panel for each spectrum, where $\theta_0$ is the angle between the incident
    ray path and the surface normal. The observations were performed on 17~July~2014~UT with a
    sub-observer longitude of 291$\odeg$W, and a Sun-Titan-Observer phase angle of 5.4\odeg, which
    is small enough that the emission angle $\mu_e$ differs from $\mu_0$ by less than 0.01.
    $\widehat{\chi^2}$ is calculated for $z_s$ in each layer of the model up to 18\,km (data points),
    and a second-order polynomial is fit to the discrete values (solid curve) to determine 
    the best fit value of $z_s$ at the minimum of the fit (grey vertical line), with the 
    1$\sigma$ uncertainties (dotted vertical lines) determined where $\widehat{\chi^2}=1$ (horizontal dashed line) in the fit.
    }
    \end{center} 
\iftwocol
    \end{figure*}
\else   
    \end{figure*}
\fi
%-----------------------------------------------------------------------------------------

\section{Test cases for retrievals}

Here we explore how our retrieval results depend on our assumptions about haze variations near the
surface and the atmospheric temperatures. We consider three scenarios for the meridional variation
in haze below 10\,km: (1) uniform haze, (2) haze opacity that increases linearly with latitude to
the north, and (3) haze opacity that is arbitrarily variable to fit the observations. In the case of
uniform haze, the aerosol optical depth is 0.22 below 10\,km at 1.5\um, whereas in the case of
variable haze, the optical depth of aerosol increases linearly from 0.03 at 40$\odeg$S to 0.10 at
80$\odeg$N. The difference in optical depths is due to the different haze structure at higher
altitudes in the two models (see Fig.~4 in \citet{A16}).  The total optical depth in both models is
$\sim$0.5 at 40$\odeg$S and increases monotonically to $\sim$0.9 at 70$\odeg$N.

We consider an arbritary haze opacity in case (3) by starting with the nominal methane profile,
fitting the spectra with a Levenberg-Marquardt optimization of the haze, and then determining the uncertainty in the methane saturation altitude as described in Appendix A. In practice, the Levenberg-Marquardt optimization fails for free parameters with discrete values, such as the saturation altitude $z_s$ in the model, so the optimization cannot fit both parameters simultaneously. By starting with a nominal methane profile for the methane optimization, we are confining our analysis to a local minimum in the spectral fitting. We do not fully consider the global parameter space in this fitting, including, for example, whether supersaturated  values of methane absorption can be balanced by significant enhancements in haze scattering. The goal of test case (3) is to evaluate our assumptions about either the uniform or linearly variable haze, and illustrate the saturation altitude sensitivity when we assume that the spectral variation is caused primarily by haze variation.

Retrievals of the saturation altitude, as well as the corresponding specific humidity and relative
humidity at the surface, are presented for the various scenarios in Fig.~\ref{f:all_humidities}. The first and third rows are also shown in Fig.~\ref{f:hum_obs}. In both of the cases where the haze is variable, the saturation altitudes and specific humidities do not show meridional variation, as is expected from the
degeneracy in opacities. The retrieved saturation altitudes and humidities for both of these cases
(bottom rows of Fig.~\ref{f:all_humidities}) are very similar to each other, and in both cases the
retrievals are still marginally consistent with saturation to the surface at very high northern
latitudes.

We also test for sensitivity to temperature by using the vertical temperature profile retrieved via
radio occultation at 80$\odeg$N during T57 by \citet{Schinder2012} for the four latitudes sampled
north of 70$\odeg$N. With cooler temperatures, the atmosphere holds less methane, so for a given
increase in opacity (and associated methane column) methane must be saturated to lower altitudes. As
illustrated in the top row of Fig.~\ref{f:all_humidities} (and in Fig.~\ref{f:hum_obs}), the saturation altitude decreases  when retrieved assuming a polar temperature profile and constant haze opacity, and the trend from $\sim$45$\odeg$N to the pole is smooth. On the other hand, with constant temperatures, the
retrievals yield a sudden increase in saturation altitudes at the four northernmost latitudes
(second row of Fig.~\ref{f:all_humidities}), which we consider unphysical. In the case of variable
haze (either arbitrarily or linearly-varying), haze opacity is independent of temperature and the choice of polar temperature profiles does not impact the retrievals.

%-----------------------------------------------------------------------------------------
%DOUBLE COLUMN FIGURE
\iftwocol
    \begin{figure*} \includegraphics[]{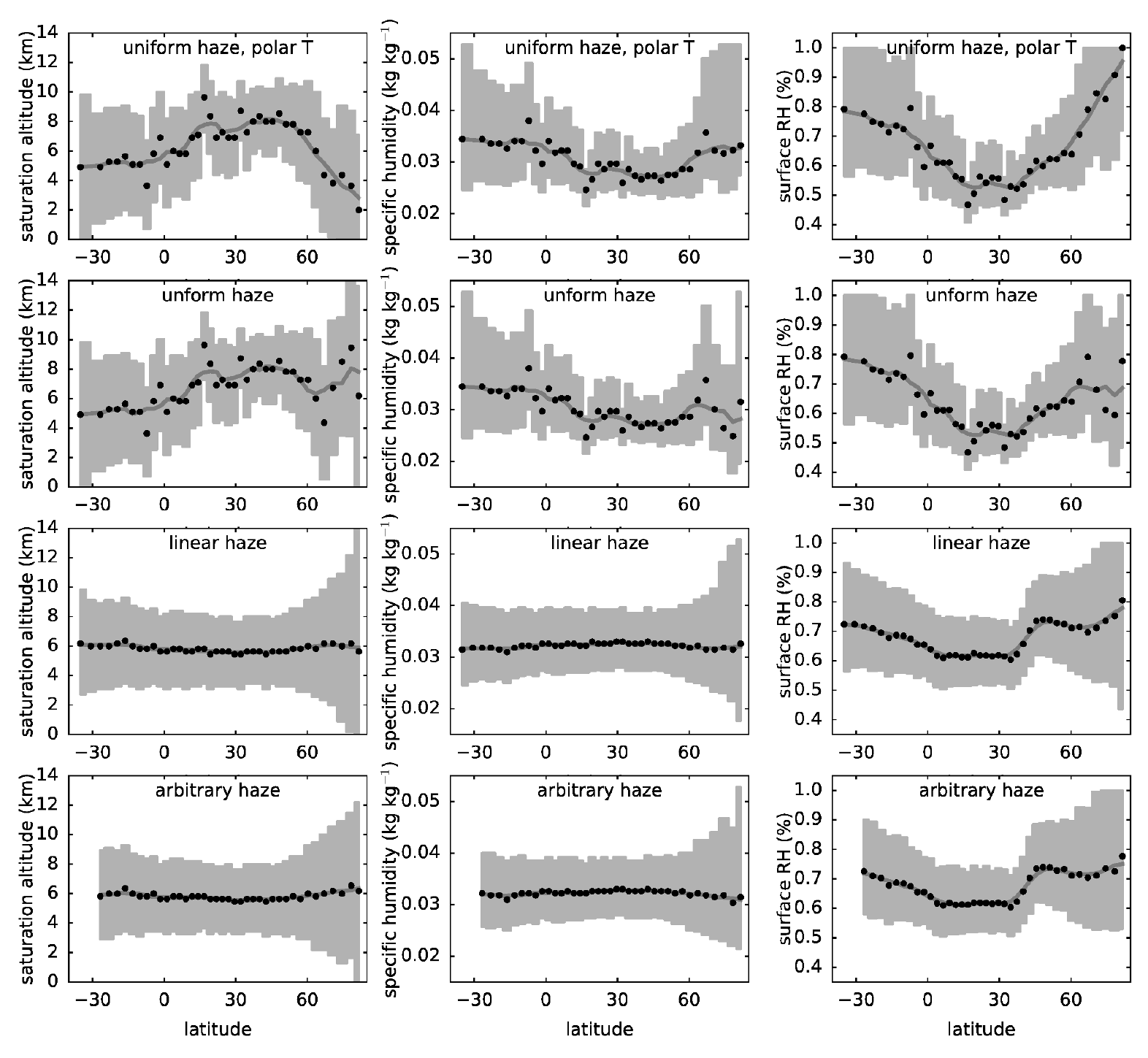} 
\else
    \begin{figure} \includegraphics[]{all_humidities_APNDX.pdf}
\fi
    \begin{center}
    \caption{\label{f:all_humidities} The meridional variation of minimum saturation altitudes (left
    column), corresponding specific humidities (middle column), and associated surface-level
    relative humidities (right column) retrieved from the observations. Retrievals assuming a
    uniform haze and a polar temperature profile (see text) are in the top row. All other retrievals
    are shown assuming the temperature profile measured by the {\it Huygens} probe at all latitudes.
    Retrievals assuming a uniform haze are shown in the second row, retrievals assuming a linear
    haze variation are shown in the third row, and models assuming arbitrary haze opacities are
    shown in the bottom row. }
    \end{center} 
\iftwocol
    \end{figure*}
\else   
    \end{figure}
\fi
%-----------------------------------------------------------------------------------------

\end{document}

%% file: aas_journal_abbrevs.tex
\def\apjl{{ApJ} }                % Astrophysical Journal, Letters
\def\icarus{{Icarus} }           % Icarus
\def\nat{{Nature} }              % Nature
\def\grl{{Geophys.~Res.~Lett.} } % Geophysics Research Letters